\documentstyle[11pt,paspconf,epsf]{article}

\def\deg   {$^\circ$}

\def\phiI  {$\varphi_i$}
\def\phiIV {$\varphi_4$}
\def\Pcos  {$\Phi^0$}
\def\Jtwo  {$J_{-21}$}

\def\Jcos  {$J_{-21}^0$}
\def\Em    {${\cal E}_m$}

\def\nuOB  {\nu_{\scriptscriptstyle {\rm OB}}}
\def\aB    {\alpha_{\scriptscriptstyle B}}
\def\nH    {n_{\scriptscriptstyle H}}
\def\NH    {\rm N_{\scriptscriptstyle H}}
\def\Ha    {${\rm H}\alpha$}
\def\Ha    {H$\alpha$}

\def\NII   {[N${\scriptstyle\rm II}$]}

\def\deg   {$^\circ$}

\def\eg    {{\it e.g.,\ }}

\def\etal  {{\it\ et al.}}
\def\intensity{\ifmmode{{\rm\ erg\ cm}^{-2}{\rm\ s}^{-1}
      {\rm\ Hz}^{-1}{\rm\ sr}^{-1}}
      \else {\ erg cm$^{-2}$ s$^{-1}$ Hz$^{-1}$ sr$^{-1}$}\fi}

\def\flux{\ifmmode{{\rm erg\ cm}^{-2}{\rm\ s}^{-1}}\else {erg
cm$^{-2}$ s$^{-1}$}\fi}

\def\fluxdensity{\ifmmode{{\rm erg\ cm^{-2}\ s^{-1}\ Hz^{-1}}}\else {erg
cm$^{-2}$ s$^{-1}$ Hz$^{-1}$}\fi}
\def\phoflux{\ifmmode{{\rm\ phot\ cm}^{-2}{\rm\ s}^{-1}}\else {\ phot
cm$^{-2}$ s$^{-1}$}\fi}
\def\eflux{\ifmmode{{\rm\ erg\ cm}^{-2}{\rm\ s}^{-1}}\else {\ erg
cm$^{-2}$ s$^{-1}$}\fi}

\def\nH {\ifmmode{\rm n_{\scriptscriptstyle H}}\else{n$_{\scriptscriptstyle H}$}\fi}
\def\NH {\ifmmode{{\rm N_{\scriptscriptstyle H}}}\else{N$_{\scriptscriptstyle H}$}\fi}
\def\Np {\ifmmode{{\rm N_{\scriptscriptstyle p}}}\else{N$_{\scriptscriptstyle p}$}\fi}

\def\ne {\ifmmode{\rm n_{\scriptscriptstyle e}}\else{n$_{\scriptscriptstyle e}$}\fi}
\def\np {\ifmmode{\rm n_{\scriptscriptstyle p}}\else{n$_{\scriptscriptstyle p}$}\fi}

\def\nHo {\ifmmode{\rm n_{\scriptscriptstyle H}^o}\else{$n_{\scriptscriptstyle H}^o$}\fi}
\def\neo {\ifmmode{\rm n_{\scriptscriptstyle e}^o}\else{$n_{\scriptscriptstyle e}^o$}\fi}

\def\RH {\ifmmode{\rm R_{\scriptscriptstyle H}}\else{R$_{\scriptscriptstyle H}$}\fi}
\def\Redge {\ifmmode{\rm R_{\scriptscriptstyle 25}}\else{R$_{\scriptscriptstyle 25}$}\fi}
\def\Fkpc  {\ifmmode{\rm F_{\scriptscriptstyle kpc}}\else{F$_{\scriptscriptstyle\rm kpc}$}\fi}

\def\emunits {\ifmmode{\rm\ cm^{-6}\ pc}\else{\ cm$^{-6}$\ pc}\fi}
\def\Em      {\ifmmode{{\cal E}_m}\else{${\cal E}_m$}\fi}
\def\aB      {\ifmmode{\alpha_{\scriptscriptstyle\rm B}}\else{$\alpha_{\scriptscriptstyle\rm B}$}\fi}

\def\nuo   {\ifmmode{\nu_o}\else{$\nu_o$}\fi}
\def\nuOB  {\ifmmode{\nu_{\scriptscriptstyle\rm OB}}\else{$\nu_{\scriptscriptstyle\rm OB}$}\fi}

\def\phiI  {\ifmmode{\varphi_i}\else{$\varphi_i$}\fi}
\def\phiIV {\ifmmode{\varphi_4}\else{$\varphi_4$}\fi}
\def\Pcos  {\ifmmode{\Phi^o}   \else{$\Phi^o$}\fi}
\def\Jtwo  {\ifmmode{J_{-21}}  \else{$J_{-21}$}\fi}
\def\Jcos  {\ifmmode{J_{-21}^o}\else{$J_{-21}^o$}\fi}

\def\sigHI {\ifmmode{\sigma_{\scriptscriptstyle\rm HI}}\else{$\sigma_{\scriptscriptstyle\rm HI}$}\fi}
\def\sigFP {\ifmmode{\sigma_{\scriptscriptstyle\rm FP}}\else{$\sigma_{\scriptscriptstyle\rm FP}$}\fi}
\def\vmax  {\ifmmode{v_{\scriptscriptstyle\rm max}}\else{$v_{\scriptscriptstyle\rm max}$}\fi}

\begin{document}

\title{Optical rotation curves beyond the HI cut-off in spirals}
\author{J. Bland-Hawthorn}
\affil{Anglo-Australian Observatory, P.O. Box 296, Epping, NSW 2121, Australia}

\begin{abstract}
In the best observed spiral galaxies, the HI is observed to decline slowly
with radius for up to 90\% of the maximum extent, before experiencing a
rapid truncation.  Early models predicted that the outer parts of spirals 
would be fully ionized by a metagalactic UV field at column depths of 
roughly 10$^{19}$ atoms cm$^{-2}$.
Weak optical line emission has now been observed beyond the HI edge in
a number of cases and, furthermore, this warm gas has been used to trace 
the dark halo far beyond the HI cut-off. But the observed emission
measures (40$-$90 mR) are higher than expected for a cosmic origin. A more
likely explanation is ionization of the warped outer disk by the blue
stellar population at smaller radius. There are many implications of
ionized warped edges, in particular, severe HI warps may be greatly
under represented in current HI surveys. We briefly discuss the 
implications of this model for the Galactic warp.
\end{abstract}


\section{Introduction}

Bochkarev \& Sunyaev (1977) first argued that the HI disks of spiral 
galaxies truncate at a few times the optical diameter because the
exponentially declining HI column eventually becomes fully ionized by
a metagalactic UV field.
After detailed calculations of the expected emission measures
(Maloney 1993; Dove \& Shull 1994), it was soon realized that these
extremely faint levels could be reached with the Fabry-Perot `staring'
technique (Bland-Hawthorn\etal\ 1994; hereafter BTVS). This method has 
since been used successfully on Sculptor galaxies (Bland-Hawthorn, Freeman 
\& Quinn 1997; hereafter BFQ), M33 and NGC 3198 (Bland-Hawthorn, Veilleux
\& Carignan 1998; hereafter BVC). In keeping with the main focus of 
this workshop, we start with a brief discussion of the dark halo in NGC 253.
But most of the article will be devoted to the source of the HI
truncation.

\begin{figure}
\label{n253}
\plotfiddle{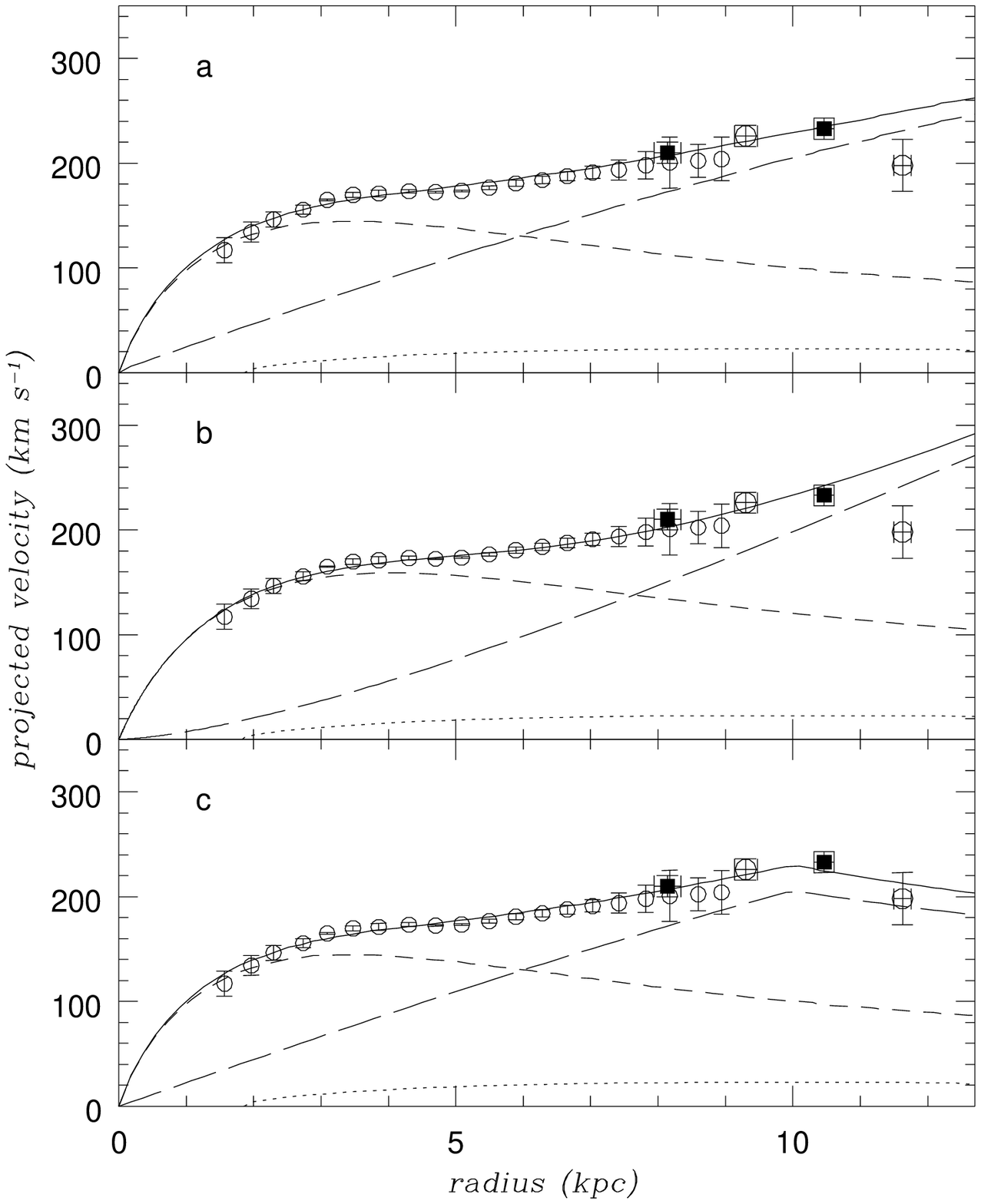}{3cm}{0}{40}{40}{-180}{-170}
\plotfiddle{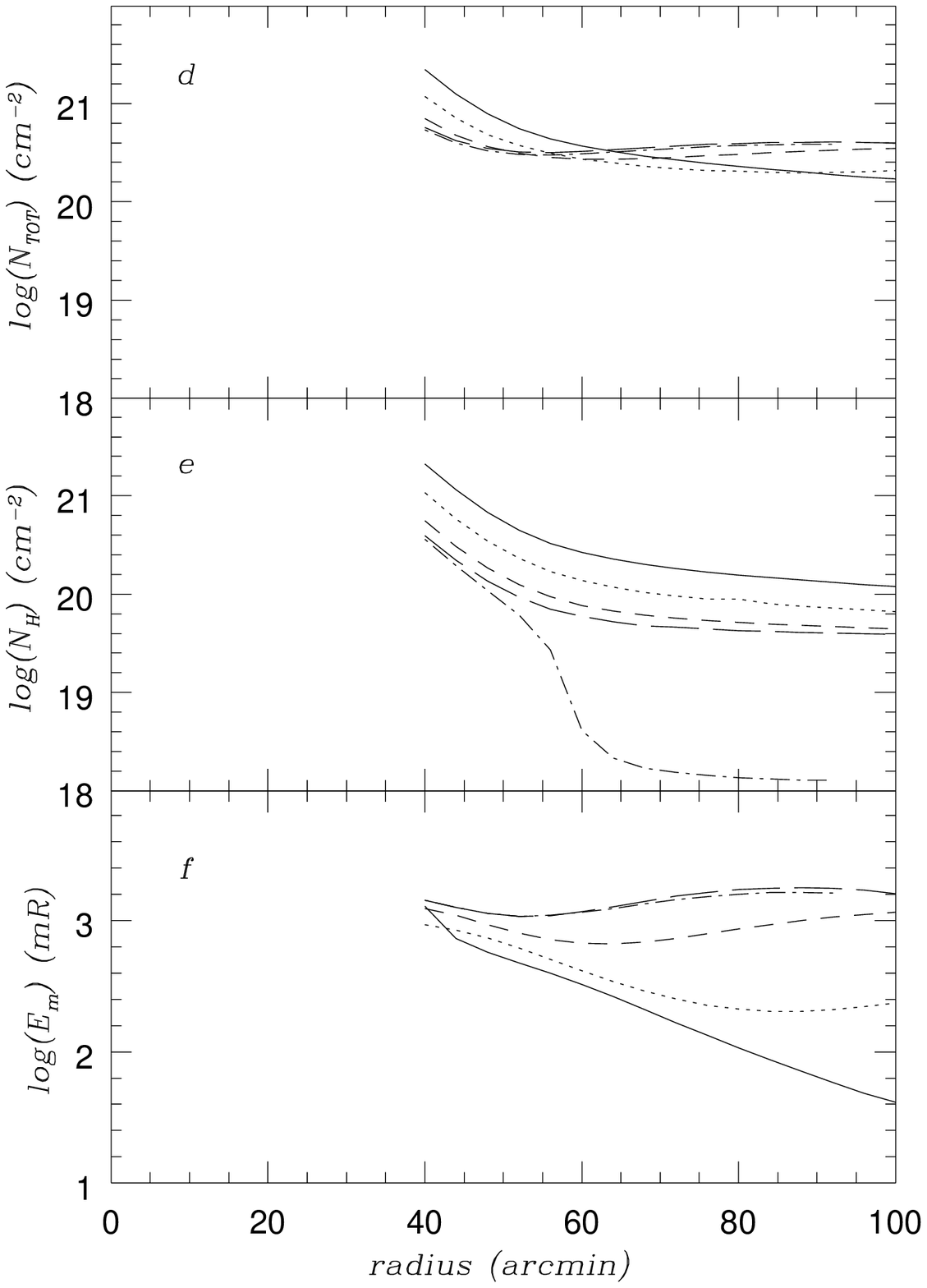}{3cm}{0}{40}{40}{10}{-73}
\caption{In the left panel ($a-c$), the HI, \Ha\ and [NII] rotation curves
are shown with representative
models for NGC 253 (see text). In the right panel ($d-f$), we illustrate 
the influence of an ionizing disk on different warp extremes in M33.
We show the projected HI, seen from the galaxy nucleus, before ($d$) and 
after ($e$) the disk ionization is switched on, and the resulting trend
($f$) in emission measure with radius. The deprojected HI data are from
Corbelli \& Schneider (1997).}
\end{figure}

\section{The dark halo of NGC 253}

In Fig. 1$a-c$, we show the kinematic measurements along the 
major axis of NGC 253 deduced from VLA HI (inner points), TAURUS-2 \Ha\ 
and \NII\ lines (outer 4 points). The HI data are from a fit to the
full velocity field (Puche, Carignan \& van Gorkom 1991; hereafter PCvG).
Whether one fits to the approaching side, receding side, or the full
velocity field, the rotation curve is still rising in the outermost
parts, a phenomenon observed in a number of low mass spirals (Puche \&
Carignan 1991).

The inflexion in the rotation curve is understood naturally to
arise from the potential of an exponential stellar disk with a much
larger dark halo (\eg Carignan \& Freeman 1985).  To illustrate this
(Fig. 1$a-c$), we show representative fits to the PCvG and TAURUS-2 
data using a 3-component
mass model (disk, halo, gas). The contribution from the HI surface
density (dotted line) is the same in all models.  In $a$, we adopt an
exponential disk (short-dashed line) and choose an isothermal sphere
for the spherical halo (long-dashed line) using the numerical approach
developed by Carignan (1985). In $b$, we have used the disk-halo
models of Dehnen \& Binney (1997).
In $c$, we have truncated the halo at a radius of 10
kpc to demonstrate a possible explanation for the last measured point.
(A detailed discussion is given in BFQ.)

The outermost TAURUS-2 \Ha\ measurement suggests
that the rotation curve may be falling beyond a radius of about 10
kpc, but there are certainly other possible explanations for its low
observed velocity (\eg tidal forces). But BFQ consider
the possibility that the rotation curve of NGC 253 is indeed falling 
beyond 10 kpc due to a truncated dark halo (see Fig. 1$c$).
This compares to the much larger dark halo distributions observed in our
Galaxy (\eg Freeman 1996) and other large spirals (\eg Zaritsky \etal\
1996).

If this interpretation is correct, there are some interesting
consequences.  The properties of dark halos are best studied in disk
galaxies for which the HI distribution extends well beyond the optical
distribution. In the outer regions of these galaxies, the ratio of
(dark matter surface density)/(HI surface density) is roughly constant
(\eg Bosma 1978; Carignan 1991). NGC 253 is not such a galaxy.  Its
HI extends only 40\% further than the optical extent. But the data
provide the first hint that where the HI
and the light are co-extensive, the dark matter also may not extend
much beyond the optical distribution. It is tentative evidence of
the apparent link between the dark matter and the HI. We emphasize
that the dark matter is still essential to generate the observed
rotation curve for NGC 253: the inferred mass ratio of dark matter to
luminous matter is about 5.

\section{What ionizes the outer edge of NGC 253?}

In NGC 253, we see ionized gas with \Em\ $\approx$ 40$-$90 mR at and
beyond the HI edge. The [NII]$\lambda 6548$/\Ha\ ratio
is close to unity which is anomalously high for diffuse gas.  

\smallskip\noindent{\bf Young stellar disk.}
BFQ attempt to explain this enhancement with recourse to standard
ionization codes and an idealised model for the young disk
(see Bland-Hawthorn \& Maloney 1997; 1998; hereafter BM). A useful 
approximation
to this model, and the predicted halo fluxes, are given in the appendix.  
Since essentially all spirals are warped, it is likely that this
interpretation has application to a large fraction of objects. P.R. Maloney
has calculated the HI truncation produced by our model in the context
of M33 (Fig. 1$d-f$). 
The curves correspond to integral sign warps with a range of maximum
amplitudes: 10\deg\ (solid), 20\deg\ (dot), 40\deg\ (short dash), 60\deg\
(long dash), modified 60\deg\ (dot-dash). The last curve is the same as
the previous curve except that the column density has been reduced by 5\%.
The existence of a truncated edge is very sensitive to the ionizing flux 
and the shape of the warp.

An important prediction of the models is that most of the low ionization
lines are greatly enhanced with respect to \Ha, and that [OIII]5007
is essentially invisible. This has largely been confirmed from
a run in October 1997 with K.C. Freeman at the MSO 2.3m using the
Double Beam Spectrograph. We obtained a total exposure time of 5.6 hours
at field positions SW3 and SW4 (BFQ). The stacked data reveal that, like [NII]/\Ha,
[SII]/\Ha\ is extraordinarily high ($\approx 0.5-0.8$) in agreement with 
the model.

\smallskip\noindent{\bf Cosmic ionizing background.}
From \Ha\ non-detections towards extragalactic HI,
the current best 2$\sigma$ upper limit (Vogel\etal\ 1995) for the
cosmic ionizing background is \Jcos\ $<$ 0.08 (\Pcos\ $< $2-4$\times 10^4$
phot cm$^{-2}$ s$^{-1}$; see appendix). This compares with \Jcos\ $\approx$ 
0.006 deduced 
from the proximity effect (Kulkarni \& Fall 1995). If the 3C 273 sight line
is truly representative, the Milky Way dominates the ionizing field out
to 400 kpc on average, and maybe 700 kpc along the polar axis. This
is an order of magnitude smaller than the scale of typical $L_*$ galaxy 
separations at the present epoch. 

BTVS show that the expected emission measure from a cloud is unlikely
to be detectable beyond 300 kpc, or maybe 500 kpc along the polar axis.
This has a number of important ramifications, \eg it is
unlikely we can detect the mutual ionization of M31 and the Galaxy. 
At this meeting, L. Blitz has laid claim to a large fraction of
high velocity clouds, placing them at extragalactic distances within
the Local Group. However, if most of these clouds produce detectable
\Ha, this interpretation is unlikely to be correct. 

\smallskip\noindent{\bf Suprathermal particle heating.} 
The enhanced low ionization lines require substantial
heating of the gas without further ionization. BFQ go
some way to achieving this with a dilute, hardened radiation
field impinging on gas with refractory element depletion. 
But the required heating rate is too low by at least a factor
of two. Many early candidates (\eg relativistic particles)
have proved unsuccessful. More promising are
subrelativistic, heavy (suprathermal) particles channelled along 
magnetic field lines to the outer disk. Their stopping lengths
can be very long so that the ionization rate is kept low, and
a single nucleus can produce many energetic ``knock on'' electrons.

\section{HI disk morphology}
There are a number of interesting consequences arising from global 
ionization of the outer disk. Here, we illustrate how different 
sources could produce quite distinct behaviour in (i) the outer
HI disk morphology, and (ii) the trend in \Em\ with radius. In the 
next section, we consider the special case of the Galactic warp.

\begin{figure}
\label{disks}
\plotone{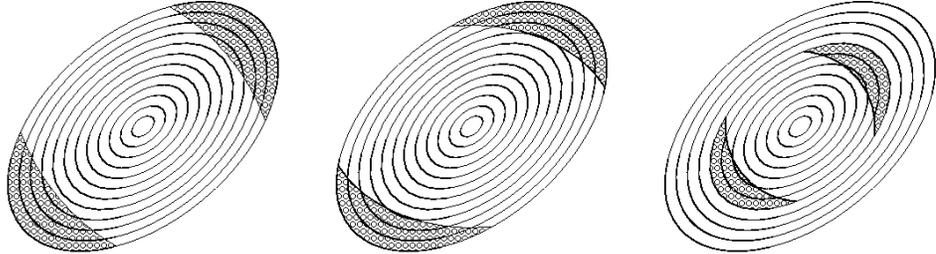}
\caption{Ionized outer regions arising from (a) a simple warp, (b) 
a precessing warp, and (c) a simple warp that folds back on itself.}
\end{figure}

In the table below,
the deprojected shape of the HI disk at a fixed column density is shown 
in the first column.  The source of the truncation and the expected trend
in \Em\ (with increasing radius) are given in the second and third
columns respectively:

{\small
\bigskip
\begin{tabbing}
{\it shape of HI disk}\hspace{0.5in} \= {\it source of truncation}\hspace{1.5in}
\= {\it radial trend} \\ \\
mirror symmetric  \> integral-sign warp $+$ internal source \> \Em\ increasing \\
axisymmetric      \> ambient radiation field                \> \Em\ decreasing \\
bisymmetric       \> tides due to external galaxy           \> \Em\ $=$ 0 \\
asymmetric        \> ram pressure from external medium      \> \Em\ complex \\
\end{tabbing}
}

The first interpretation assumes a simple warp for the HI, a characteristic 
of most warps (Briggs 1990), although one can certainly construct different 
scenarios (see Fig. 2) for ionization by a central source.  The prediction 
of \Em\ increasing is confirmed in Fig. 1$f$ although only in the limit of
strong warps ($>30$\deg).  These models predict azimuthal incompleteness in 
the HI distribution at critical velocities.  Careful kinematic analysis of 
the best observed spirals could put strong limits on each of these models. 
The second interpretation is
the conventional picture in terms of a metagalactic UV field (see $\S1$)
but, to date, we see no evidence for such a model. The third and fourth
models are speculative at best: it is difficult to see how either could
produce well defined HI truncations.

In order to explore these different cases, we are currently studying a 
range of galaxy types in different environments, with the aim of tracing
the dark matter beyond the HI limit.  The galaxies chosen for
detailed study include M31, M33, M81, M83, NGC 628, NGC 3198, NGC 5266,
Fourcade-Figueroa, and the Sculptor Group.  The most detailed work has
concentrated on NGC 253 (BFQ), M33 and NGC 3198 (BVC).

\section{The Galaxy}
In Fig. 3, we illustrate the predicted halo ionizing field.
The 21 cm data are from Burton (1984) and Burton \& te Lintel Hekkert (1986).
This is normalized to explain the Magellanic \Ha\ stream detections
(Weiner \& Williams 1996). The northern HI warp is easily truncated
by the disk radiation; we suggest that the southern warp is much less 
pronounced due to the ionizing influence of both the LMC and the 
inner disk. 

The predicted levels of \Ha\ emission should be easily detectable 
(150$-$200 mR) with WHAM, and only marginally by other all-sky \Ha\ 
surveys. We anticipate that the projected \Ha\ emission off the plane will 
show peaks at $l=70$\deg\ and $l=250$\deg\ respectively, with the southern 
extension being more pronounced.  It should be possible 
to extend the Galactic rotation curve beyond the current HI limit. 
More generally, it seems highly likely that severe warps in spirals,
particularly those with UV-bright inner disks, are suppressed by disk 
ionization.

\begin{figure}
\label{halo}
\plotone{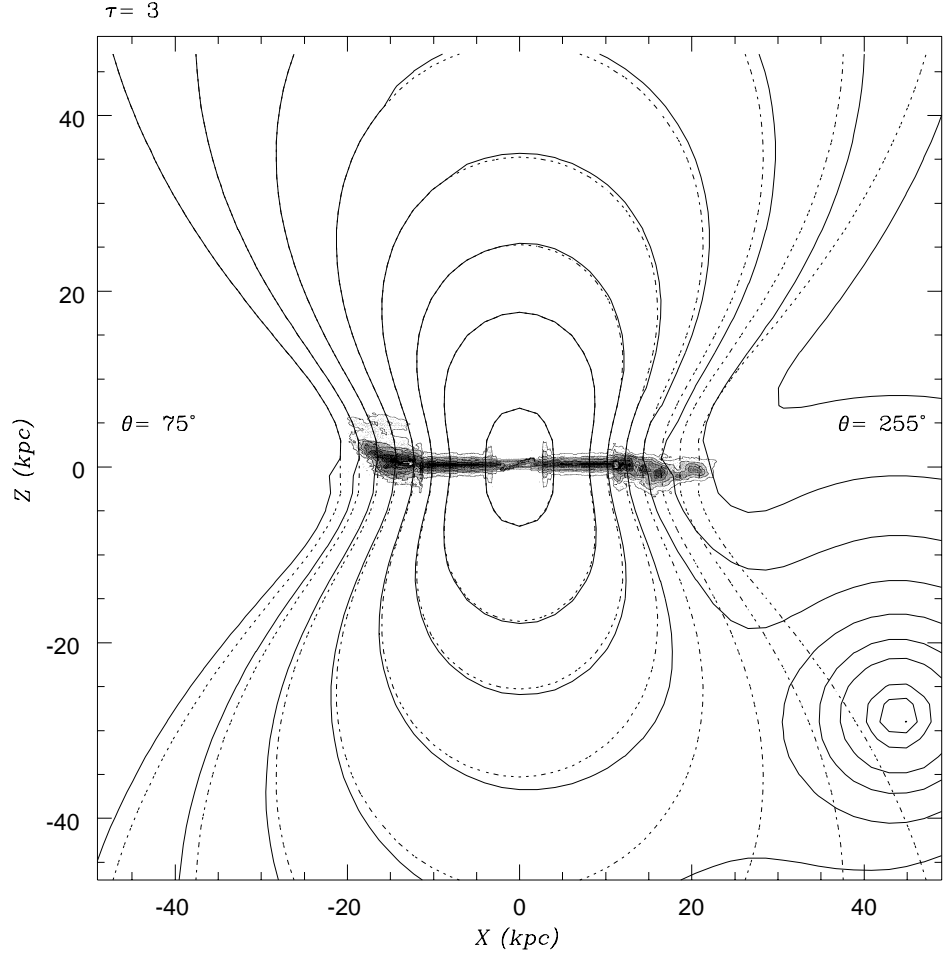}
\caption{The galactic halo ionizing field. The coordinates are
with respect to a plane perpendicular to the Galactic disk (with
the Galactic Centre at the origin) at a constant galactic azimuth
angle (75\deg, 255\deg). The dotted lines show the ionizing flux 
($\varphi_4$) due to the galactic disk; the solid lines include the
contribution from the LMC. The opacity of the HI disk (shown in 
half tone) has been included. The contours, from outside in, are
for $\log \varphi_4 = 1, 1.25, 1.5, 1.75, 2, 2.25, 2.5, 3$ phot cm$^{-2}$ 
s$^{-1}$. The minor contribution from the Galactic corona is omitted (BM).}
\end{figure}

\section{Discussion}

The ionization models for warped edges suggest that essentially any
spiral with a young, stellar population can be reasonably traced to
and beyond the HI limit.  In the months to come, there will be half a 
dozen spirals with optical rotation curves at and beyond the HI edge. 
For these objects, the total mass in H$^+$ is comparable to the HI mass.

It is possible to use the `staring' method down 
to surface brightness levels at least 6 magnitudes below sky in a 1\AA\ 
band (BTVS). It may be possible to extend this limit further
with telescope nodding, charge shuffling and deep depletion CCD arrays,
all of which will be in place at the AAT by early 1998. {\it In principle,} 
stellar rotation curves can be traced to several magnitudes below sky,
although the Lorentzian instrumental profile is a handicap to reaching
much beyond a Holmberg radius! There are good reasons for pushing the 
stellar rotation curves. In special cases, the H$^+$ kinematics may need 
second order corrections (\eg asymmetric drift) for ionization effects 
(\eg Yorke, Bodenheimer \& Tenorio-Tagle 1982; Yorke \& Welz 1996).

We stress that 21 cm mapping of spirals is crucial to this work, and is 
substantially more efficient than `staring' over the HI extent.  (For 
reasons outlined in $\S4$, it is even worth the effort to revisit archived 
21 cm data.)  It takes 0.5$-$1 nights to obtain a pair or triplet of optical
kinematic measurements. More rapid progress is possible with
a dedicated 2.5m$-$5m telescope (see BFQ, eqn. 7).
These measurements are tied to the HI data and must be corrected for warping 
(and flaring) of the disk. While Fabry-Perot `staring' has extraordinary 
potential for many problems, \eg R.J. Reynolds' outstanding 
work with the WHAM instrument, it merely complements the deepest 
measurements obtainable in other bands.

%*********************************************************************
\acknowledgments
I am indebted to K.C. Freeman, P.R. Maloney, H.W. Yorke, W.B. Burton, and 
M.A. Dopita for assistance with several aspects of this work.

%*********************************************************************

%*********************************************************************
\appendix
\noindent\section{Ionization models}

We present a useful approximation to the Galactic halo ionizing field
deduced by BM.  We assume an electron temperature T$_e \simeq 10^4$K, 
as expected for gas photoionized by stellar sources,
for which the Case B hydrogen recombination coefficient is $\aB \simeq
2.6 \times 10^{-13} (10^4/T_e)^{0.75}$ cm$^3$ s$^{-1}$. At these
temperatures, collisional ionization processes are negligible. In this
case, the column recombination rate in equilibrium must equal the
normally incident ionizing photon flux, $\aB n_e N_{H^+} = \varphi_i$,
where \phiI\ is the rate at which Lyc photons arrive at the cloud
surface (photons cm$^{-2}$ s$^{-1}$), $n_e$ is the electron density
and $N_{H^+}$ is the column density of ionized hydrogen. The emission
measure is ${\cal E}_m = \int n_e n_{H^+}\;dl =n_e n_{H^+} L\ 
{\rm cm^{-6}\; pc}$ where $L$ is the thickness of the ionized region.
The resulting emission measure for an ionizing flux \phiI\ is then
${\cal E}_m = 1.25\times 10^{-2} \varphi_4 \ {\rm cm^{-6}\; pc}$ 
($= 4.5 \varphi_4 \ {\rm mR}$) where
$\varphi_i = 10^4 \varphi_4$.  For an optically thin cloud in an
isotropic radiation field, the solid angle from which radiation is
received is $\Omega = 4\pi$, while for one-sided illumination,
$\Omega=2\pi$.  For our disk model, however, $J_\nu$ is anisotropic and 
$\Omega$ can be considerably less than $2\pi$.  

For the cosmic field, \Jcos\ is the ionizing flux density of the
cosmic background at the Lyman limit in units of 10$^{-21}$ erg
cm$^{-2}$ s$^{-1}$ Hz$^{-1}$ sr$^{-1}$; \Pcos\ ($=\pi$\Jcos$/h$) is the
equivalent photon flux at face of a uniform, optically thick slab.

For the Galactic halo (Fig. 3), to a good approximation, the poloidal UV 
radiation field is
\begin{equation}
\varphi_4 = 2.8\times 10^6 \ e^{-\tau}\ {\rm r_{kpc}}^{-2} \cos^{0.6\tau+0.5} \Theta \ \ \ \ \ \ \ \ {\rm phot\ cm^{-2}\ s^{-1}}
\end{equation}
where $\Theta$ is the polar angle ($0 \leq \Theta < \frac{\pi}{2}$) and the 
Lyman limit optical depth 
$\tau \leq 10$. It follows that the solid-angle averaged flux is
\begin{equation}
\bar{\varphi_4} = {{2.8\times 10^6\ e^{-\tau}}\over{(0.6\tau+1.5) {\rm r_{kpc}}^2}} \ \ \ \ \ \ \ \ {\rm phot\ cm^{-2}\ s^{-1}}.
\end{equation}
In order to explain Magellanic Stream \Ha\ detections, our model favours 
$\tau \approx 2.8$.
\end{document}